\documentstyle[eqsecnum,aps,prb,twocolumn]{revtex}

\begin{document}
\draft
\preprint{HEP/123-qed}
\title{Absence of a structural transition up to 40GPa in MgB$_{2}$ and the relevance of magnesium non-stoichiometry}

\author{P. Bordet}
\address{Laboratoire de Cristallographie, CNRS, B.P. 166, 38042 Grenoble Cedex 9, France}
\author{M. Mezouar}
\address{European Synchroton Radiation Facility, B.P. 220, 38043 Grenoble Cedex 9, France}
\author{M.  N$\acute{u}\tilde{n}$ez-Regueiro$^{*}$ and M. Monteverde$^{**}$}
\address{Centre de Recherches sur les Très Basses Températures, CNRS, B.P. 166, 38042 Grenoble, Cedex 9, France}
\author{M. D.  N$\acute{u}\tilde{n}$ez-Regueiro}
\address{Grenoble High Magnetic Field Laboratory, MPI-FKF and CNRS, B.P. 166, 38042 Grenoble, Cedex 9, France}
\author{N. Rogado, K. A. Regan, M. A. Hayward, T. He, S. M. Loureiro, and R. J. Cava}
\address{Department of Chemistry and Materials Institute, Princeton University, Princeton,
NJ 08544, USA}

\date{March 9, 2001}
\maketitle
\begin{abstract}
 We report measurements on MgB$_{2}$ up to $\sim$40GPa.
Increasing pressure yields a monotonous decrease of the lattice parameters and of the c/a
ratio, but no structural transition down to parameters smaller than those of AlB$_{2}$.
The transition superconducting temperature also decreases with temperature in a sample dependent way.
The results are explained by an increase of the filling of the 2D $p_{xy}$ bands
with pressure, the Mg stoichiometry determining the starting position of the Fermi level. Our measurements indicate that these hole bands are
the relevant ones for superconductivity.
\end{abstract}

\narrowtext
    The recent discovery \cite{Akimitsu} of superconductivity in MgB$_{2}$ at $\sim$40K gives an
unprecedented opportunity to study a high temperature superconductor different from
the cuprates. Although up to now, most of the results point towards a conventional microscopic
origin, certain characteristics of the band structure \cite{Kortus,Belashchenko,Satta,An}
suggest similarities with the cuprates. Magnesium atoms yield both electrons to the boron atoms,
so we are in a situation similar to that of graphite, where the $p_{z}$ orbitals give rise to the $\pi$
bands responsible for its transport properties. However, in MgB$_{2}$ the magnesium
ions also lower the energy of these $\pi$ bands, allowing the $\sigma$ bands formed
by the $p_{xy}$ orbitals to cross the Fermi surface, furnishing holes that have a strong
two-dimensional (2D) character. These holes have an unusually strong coupling to the boron stretching
bonds yielding an intermediate superconducting coupling strength  \cite{Kong}.
Besides, it has been shown that Al doping has drastic consequences on the structure
and behaviour of this material \cite{Slusky}: the reduction of the c lattice parameter
from 3.5Å down to 3.4Å  causes the disappearance of superconductivity. This suggests that the
proximity of MgB$_{2}$ to a structural instability could be one of the factors enhancing the
superconducting transition temperature ($T_{C}$). Another possibility has been evoked from band
structure calculations: the filling of the $p_{xy}$ bands through electron transfer from the $Al$
atoms going against superconductivity \cite{An}.
As stated in Ref. \cite{Jorgensen}  one of the key parameters to test this hypothesis is obviously pressure.
A monotonic dependence of the structure with pressure has been previously reported,
measurements going only up to 8GPa \cite{Vogt,Prassides}.
On the other hand, the pressure results concerning the superconducting transition reported up
to date on MgB$_{2}$ clearly disagree. High pressure quasi-hydrostatic resistivity measurements
($\leq$25GPa) \cite{Monteverde} reported a parabolic or linear decrease of $T_{C}$ with pressure.
While hydrostatic low pressure measurements ($\leq$2GPa) showed a linear dependence
of $T_{C}$ at rates of -1.6K/GPa \cite{Lorenz} or -2K/GPa \cite{Saito}, considerably steeper than
that of the linear dependent samples of the high pressure measurements, $\sim$-0.8K/GPa
 \cite{Monteverde}.\\
Here we  perform measurements of $T_{C}$ and of structure parameters for MgB$_{2}$ for pressures up to near 40GP.
We correlate our results to the evolution of the electronic structure.
We illustrate this point by a tight-binding calculation of the pressure dependence of the relevant bands.
We can conclude that there is no structural transition, the decrease of $T_{C}$ with increasing
pressure being due to the transfer of electrons from the 3D $\pi$ bands to the 2D $\sigma$ bands.
Within this picture, the discrepancies in the $T_{C}$ dependence between different samples
can be attributed to Mg non-stoichiometry.\\
 Magnesium diboride has the AlB$_{2}$-type hexagonal structure, with a=3.08Å,
c=3.51Å, space group P6/mmm, Z=1 \cite{Jones}. The atomic positions are Mg at 1a (0 0 0), B at
2d (1/3 2/3 1/2) (inset of Fig. 1). The structural arrangement can be described as the alternate
stacking of planes of boron atoms forming a honeycomb lattice, and planes of
magnesium atoms forming a triangular one. Each Mg atom is at the center of an
hexagonal prism made of boron atoms at a distance of $\approx$2.5Å. Each boron atom is
surrounded by three other boron atoms forming an equilateral triangle at a distance of
a/$\sqrt{3}\approx$1.78Å, while the in-plane Mg-Mg distances are equal to the a parameter.\\
The sample employed were synthesized by direct reaction of the
elements. Starting materials were bright magnesium flakes (Aldrich Chemical,
Milwaukee, Wisconsin) and sub-micrometer amorphous boron powder (Callery Chemical, Evans
City, Pennsylvania), lightly mixed in a half-gram batch, and
sealed in a molybdenum tube under argon. This tube was in turn sealed in
an evacuated quartz ampoule. The material was heated 1 hour at 600$^{\circ}$C, 1 hour at
800$^{\circ}$C, and 2 hours at 950$^{\circ}$C, and then lightly ground, to provide sample A.
The same powder was then hot-pressed at 10Kbar 1 hour at 700$^{\circ}$C to supply the ceramic sample B.
The preparation of sample C was identical but kept only 1 hour at 950$^{\circ}$C;
while sample D was prepared using a tantalum tube heated 3 hoursat 900$^{\circ}$C.
Sample E was heated 1 hour in an Ar/H$_{2}$ flow.\\
    The electrical resistivity measurements were performed in a sintered diamond
Bridgman reinforced anvil apparatus using a pyrophillite gasket and two steatite disks as the
pressure medium \cite{Wittig}. The Cu-Be device that locked the anvils can be cycled between 1.2K
and 300K in a sealed dewar. Pressure was calibrated against the various phase
transitions of Bi under pressure at room temperature, and by superconducting Pb and Bi
manometers at low temperature. The overall uncertainty in the quasi-hydrostatic
pressure is estimated to be $\pm$15\%. The pressure spread across the sintered diamond
anvils was previously determined on Pb-manometers to be of about 1.5-2GPa depending
on the applied pressure. The temperature was determined using a calibrated cernox
thermometer with a maximum uncertainty (due mainly to temperature gradients across
the Cu-Be clamp) of 0.5K. Four probe electrical resistivity d.c. measurements were
made using a Keithley 2182 nanovoltmeter combined with a Keithley 220 current
source and by using platinum wires to make contact to the sample.\\
The equation of state of MgB$_{2}$ was determined up to $\approx$39GPa by angle-resolved
X-ray diffraction at the high pressure beamline ID30 of the European Synchrotron
Radiation Facility. The MgB$_{2}$ powder from sample A was loaded into a membrane-
driven diamond anvil cell with diamond tips of diameter 300$\mu$m and stainless-steel
gasket with hole of 120$\mu$m. Nitrogen was used as pressure transmitting medium in order
to keep good hydrostatic conditions. The pressure was determined by the ruby
fluorescence method \cite{Mao} with a precision of 0.1 GPa. X-ray powder diffraction patterns
were recorded every $\approx$3GPa by angle-resolved X-ray diffraction using focused
monochromatic beam at wavelength $\lambda$=0.3738Å. The X-ray signal was averaged over
the whole sample area with a 30mm diameter pinhole. The diffraction patterns were
recorded on Mar345 image plate detector located at 360mm from the sample. They
were analyzed using the software package Fit2D \cite{Hammersley}. The sample to detector distance
and the image plate inclination angles were precisely calibrated using a silicon standard
located at the sample position. After removal of spurious peaks, the corrected images
were averaged over 360$^{\circ}$ about the direct beam position, yielding Intensity vs.
2$\theta$ diffractograms. These data were analyzed by the Rietveld technique using the
Fullprof \cite{Rodriguez}  software. The peak shape was modeled with a pseudo-Voigt function.
Cell and profile parameters and an overall thermal parameter were refined at each pressure.
On increasing pressure, phases of solid molecular nitrogen also appeared \cite{Hanfland}.
These phases were taken into account by the cell constraint refinement technique.
The data could be successfully fitted with the AlB$_{2}$-type structure up to the highest
pressure investigated $\approx$39GPa. The variable overlap of the MgB$_{2}$ Bragg peaks
with those from the different phases of molecular N could lead to some inaccuracy in the
determination  of the MgB$_{2}$ cell parameters under pressure. Therefore, the
MgB$_{2}$ cell parameters were also directly determined from the angular positions of the
strong and well defined reflections (110) and (101). The results found by
both methods were in excellent agreement.\\
Fig. 1 shows the observed p(V) dependence; the line is the fit using Vinet's equation of state \cite{Vinet}:
\[p=3B_{0}\frac{(1-f_{\nu})}{f_{\nu}^{2}}exp[\frac{3}{2}(B'_{0}-1)(1-f_{\nu})]\]
where $f_{\nu}=(V/V_{0})^{1/3}$. The variable of the least squares fit were $B_{0}$, $B'_{0}$ and $V_{0}$,
representing the bulk modulus, its derivative and the cell volume at room pressure,
respectively. The obtained values were $B_{0}$=150(5)GPa, $B'_{0}$=4.0(3) and $V_{0}$=29.00(4)Å$^{3}$
($V_{0}$ is in agreement with the 28.99(1)Å$^{3}$ value reported in Ref. \cite{Jones}).
The pressure variation of the lattice parameters and and c/a are shown in Fig.2.
Both cell parameters decrease monotonically with increasing pressure. No sign of structural transition is seen.
It is worth noting that the Rietveld refinements did not indicate the presence of a structural change
for MgB$_{2}$ in the whole pressure range investigated. The c/a ratio decreases linearly up to
38.9GPa, with a slope of -1.3 10$^{-3}$ GPa$^{-1}$. As predicted theoretically by Loa and Syassen\cite{Loa},
the compression is more isotropic than expected for a layered system.\\
Fig. 3 shows the evolution of $T_{C}$ with pressure for five samples of MgB$_{2}$. We add
for comparison, the rates reported by Lorenz et al. \cite{Lorenz} and Saito et al. \cite{Saito},
-1.6K/GPa and -2K/GPa, respectively. Clearly they give different behaviours:
samples A, B and E follow a quadratic dependence starting with a small slope of -0.35K/GPa,
while samples C and D have a linear decreasing rate but with a weaker slope than those obtained
in the quoted references. It is true that those measurements have been done in hydrostatic apparata (<2GPa) while ours are performed in a quasihydrostatic anvil
system. However, the spread even among our samples shows that there is an intrinsic
reason for those differences.\\
    The differences in the behavior of the resistance of our samples correlates
with the $T_{C}$ pressure dependencies. The samples that are metallic at all temperatures and
pressures have a steeper $T_{C}$ pressure dependence than those that display upturns in the
resistivity at high pressures. It can be argued that the temperature dependence of the
resistivity may be affected by the powder nature of the sample. However, four of our five
samples are powder samples and only half of them are totally metallic.\\
    The important point is that the $T_{C}$ pressure dependence normally constitutes an
intrinsic parameter, not affected by the powder, ceramic or monocrystalline nature
of the sample. An example is the case of high temperature superconducting cuprates \cite{Nunez},
where the differences in $T_{C}$ pressure dependencies among different groups, have been tracked
down to intrinsic effects, such as oxygen ordering \cite{Sadewasser,Schilling}, or doping \cite{Wij,Taka}.
The observed differences for MgB$_{2}$ should then not be neglected as spurious results,
but a search for the hidden parameter explaining them in a coherent manner, is necessary.\\
We would like to point out that the issue of non-stoichiometry, that has been neglected up to now,
should be crucial in this material. The control of the Mg concentration is difficult at the
temperatures at which the samples are synthesized. The band structure is complex, with different
bands crossing the Fermi level. Mg defects will alter the occupancy of these bands,
besides being scattering centers for off plane carriers.\\
In Fig.4 we simulate the pressure dependence of these bands by adapting the
tight-binding model used for graphite \cite{Wallace}, and fitting it to the MgB$_{2}$
electronic structure obtained from first principles \cite{Kortus}.
Details will be given elsewhere. As in Ref. \cite{Belashchenko} the hopping parameters are scaled with
the inverse of the square of the lattice parameter variation. If we consider that the compression
of c is stronger than the one of a (experimentally the c/a ratio has decreased 4\% at 35Gpa),
and thus we take the last one as constant, only the 3D $p_{z}$
bands will change with pressure. Therefore the most important variation will be their
shift to higher energies. The consequence of this will be the electron transfer of electrons
from these $\pi$ bands to the $p_{xy}(\sigma$) bands. The decrease of $T_{C}$ with pressure can
then be attributed to this filling, these 2D bands resulting the relevant ones for
superconductivity.\\
According to Fig. 2 of Ref. \cite{An}, for the stoichiometric material the Fermi level
falls just on the verge of a steep decrease of the density of states of these hole bands. A
shift of the Fermi level to higher energies will cause a decrease of the density of states of these carriers,
and the observed decrease of $T_{C}$. Though here exact Mg concentration could not be determined
due to very small sample size (quantitative work on this matter is in progress), we can expect
that with Mg deficiency the Fermi level will be in the flatter region of the 2D band. In this case pressure will not produce a
strong effect (samples A, B and D). Instead for more stoichiometric samples (samples C, D and
the data from Refs. \cite{Lorenz} and \cite{Saito}), the compression induced shift will cause a more dramatic decrease
of $T_{C}$.\\
    In conclusion we have shown that the lattice parameters of MgB$_{2}$ monotonously decrease
with pressure up to 40GPa. Although we have attained smaller lattice parameters than for AlB$_{2}$,
we do not observe a crystallographic transition. The pressure dependence of $T_{C}$ can be
naturally explained by considering that the 2D $p_{xy}$ holes are the driving carriers for
superconductivity. Furthermore, the structural instability that has been found on Al substitution
is also probably the result of the complete filling of the $p_{xy}$ bands, that can be expected in
MgB$_{2}$ at even higher pressures. Finally, the system should be extremely sensitive to
Mg non-stoichiometry; finding a way to control this non-stoichiometry may be the simplest
way to higher $T_{C}$.\\

\narrowtext

\begin{figure}
\caption{Absolute and relative volume pressure dependence of MgB$_{2}$.
Inset: structure of MgB$_{2}$; small (big) spheres are boron (magnesium) atoms.
The coordination polyhedron of Mg atoms is shadowed.}
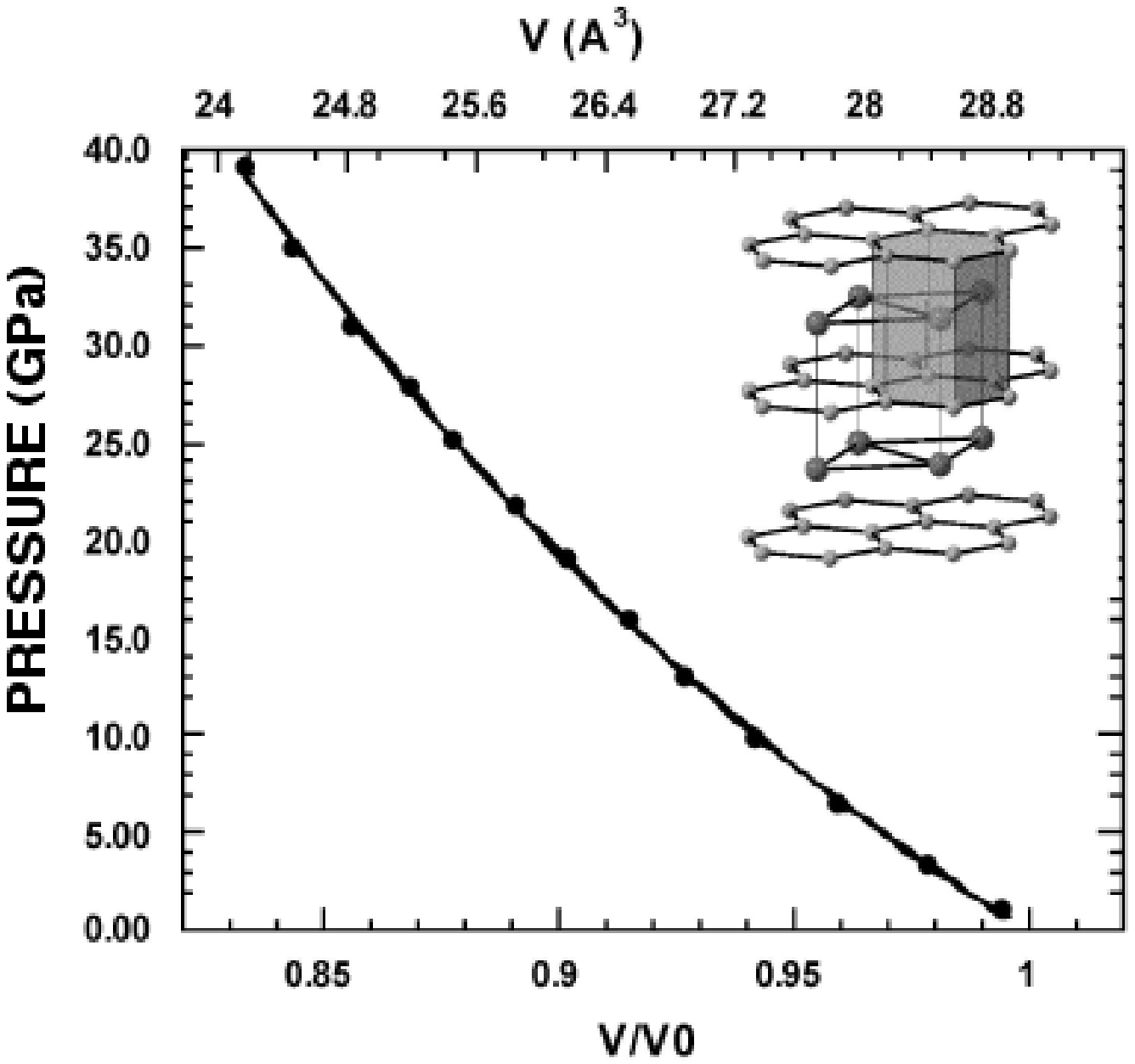
\label{BordetFig1bis.eps}
\end{figure}

\begin{figure}
\caption{Pressure dependence of the lattice  parameters of MgB$_{2}$.}
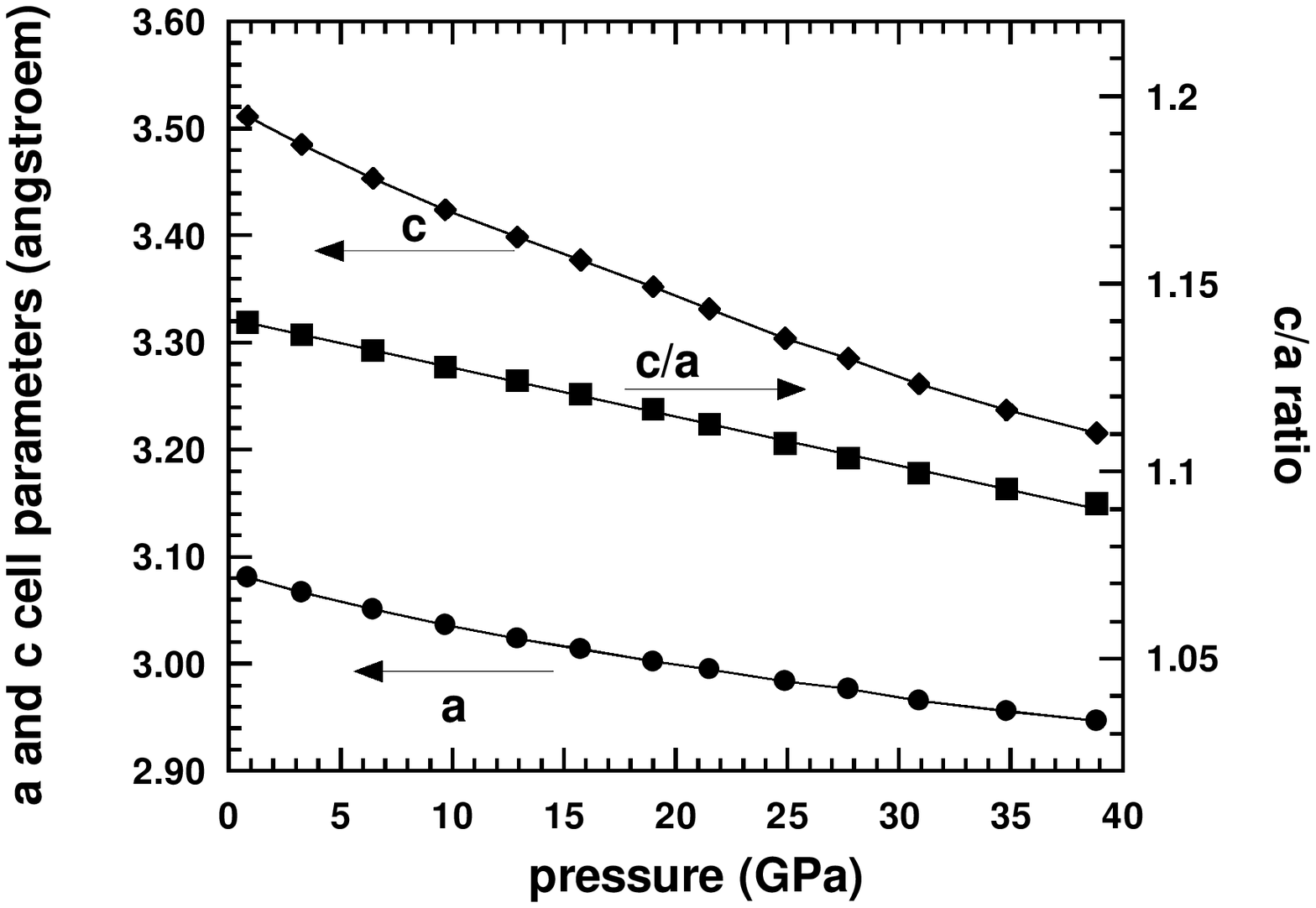
\label{BordetFig2.eps}
\end{figure}

\begin{figure}
\caption{Pressure dependence of the superconducting transition temperature of MgB$_{2}$.
Circles: sample A; squares: sample B; triangles: sample C; inverted triangles: sample D;
diamonds: sample E. The solid (dashed) line is the slope reported in Ref. 12 (13).}
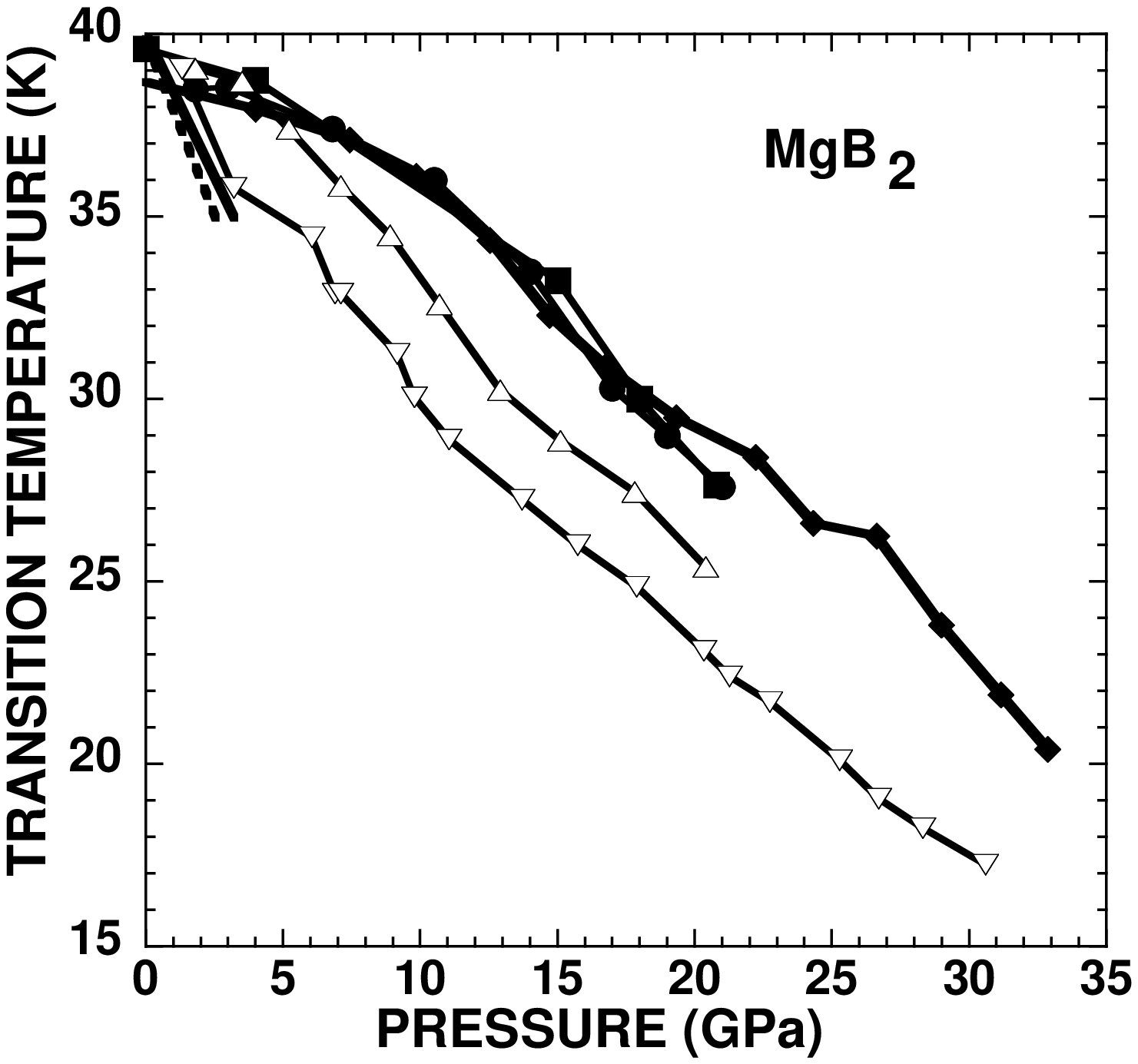
\label{BordetFig3.eps}
\end{figure}

\begin{figure}
\caption{Tight-binding calculation for the relevant bands of MgB$_{2}$ around the Fermi level.
The dashed line corresponds to c-compression (35GPa).}
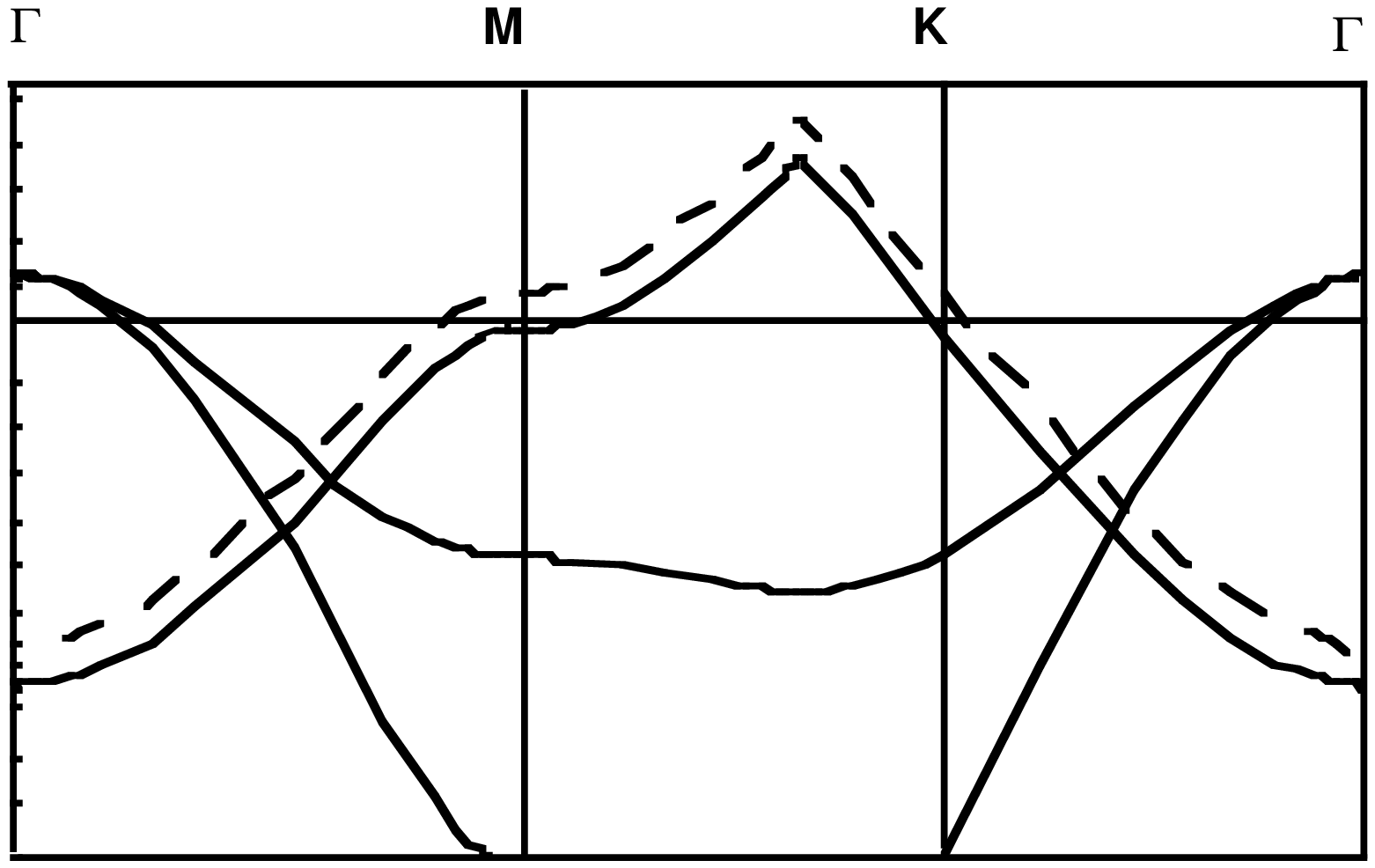
\label{BordetFig4.eps}
\end{figure}

\
\narrowtext
\end{document}